\documentclass[aps,prc]{revtex4}
\usepackage{graphicx}
\usepackage{dcolumn}
\usepackage{bm}
\newcommand{\PbPb}{\ensuremath{\mbox{Pb--Pb}}}
\newcommand{\sqrtSnn}{\ensuremath{\sqrt{s_{\mathrm{NN}}}}}
\newcommand{\pt}{\ensuremath{p_{\mathrm{T}}}}
\newcommand{\fq}{\ensuremath{F_{\mathrm{q}}}}
\newcommand{\fqe}{\ensuremath{F_{q}^{e}}}
\newcommand{\cpq}{\ensuremath{C_{p,q}}}
\begin{document}
\title{Evidence on the absence of critical transition in AMPT \\  for $\PbPb$ collisions at $\sqrtSnn$ = 2.76 TeV}

\author{Ramni Gupta, Rohni Sharma \\Department of Physics $\&$ Electronics, University of Jammu, Jammu (J\&K), India \\\ email: $ramni.gupta16@gmail.com$}
%
\date{\today}
\begin{abstract}
Event-by-event fluctuations in the spatial patterns in charged particles generated in Pb--Pb collisions at the center-of-mass energy $\sqrtSnn=2.76$ TeV are studied within A MultiPhase Transport (AMPT) model.  The spatial patterns of the particles generated in the ($\eta, \phi$) space for $|\eta| \le 0.8$ are studied  using the methodology of intermittency and erraticity analysis.  We find negative intermittency for charged particles generated in a range of $\pt$ windows.  This result contrasts sharply from what is expected for a quark-gluon plasma undergoing hadronization by a second-order phase transition. Appropriate scaling behavior is examined, resulting in definitive scaling exponent $\nu_{-}$. Event-by-event fluctuations in the spatial patterns quantified by an index, named erraticity index are determined for different $\pt$ bins $\leq 1$ GeV/c, for AMPT model. This is the first time that the intermittency and erraticity indices are determined for any model at such high energies. The results presented here  can be used for comparison with the fluctuation properties of the experimental data and hence can help the development of a wider scope of understanding of  validity of the particle production process by AMPT at these energies on the one hand,  and of the true nature of the real data on the other.
\end{abstract}
\maketitle
\section{Introduction}
Heavy ion collisions at RHIC~\cite{ref1a:rhic,ref1b:rhic,ref1c:rhic,ref1d:rhic} and recently at LHC~\cite{ref2a:lhc,ref2b:lhc} have revealed that
a state of matter is created which comprises of strongly interacting quarks and gluons commonly now known as sQGP ~\cite{ref3a:sqgp,ref3b:sqgp}. No comprehensive theoretical model exists which can address all the complexities of the physics involved in 
these relativistic heavy-ion collisions. Different treatments and techniques are applied to various aspects of the collision process, each with its own set of assumptions and parameters to describe the system created by these collisions. From the final state of hadrons that emerge in these collisions, one can extract relevant information about different properties of the dense medium that is created. 
Global observables, such as multiplicity distributions, describe the behavior of the system as a whole in contrast to local observables, such as high-\pt\ jets. A successful model focused on one aspect of the problem may not say much about other aspects, but should at least not contradict what is observed. Our study here is an initial attempt to understand the detailed nature of the global properties as manifested in local fluctuations. We use simulated events from a well-known event generator as a tool to develop our method of analysis so that at a later stage the same method can be applied to the real data.
 \\

\par
Phase diagram of the strongly interacting matter is still not understood completely. Widely acceptable school of thought, based on the lattice QCD calculations and the study of experimental data, believes that the QGP at low baryon chemical potential cools and forms hadrons in a continuous cross over manner and expects a first order phase transition for $\mu_{B}$ above a critical point. Thus, among the various properties of the dense matter created in heavy-ion collisions that have been of interest to investigators is the nature of phase transition (PT), both in the initial stage from hadrons to quarks and in the final stage from quarks back to hadrons \cite{ref:b} and to locate the critical point. 
A fair amount of effort has been devoted to the former problem of formation in the Beam Energy Scan (BES) program at low energies. Less attention has been given to the latter problem of hadronization because of complications arising from various factors related to fluctuations in particle momenta and their multiplicity in each event. It is known in statistical physics that a fundamental characteristic of the critical behavior of a system undergoing PT is that it exhibits fluctuations of all scales. Thus to find signals of quark-hadron PT and thus learn whether the system has underwent  critical behavior, one should look for clustering of produced particles of various sizes.\\

\par
In dealing with the complications associated with fluctuations, scaling properties of factorial moments (to be described late) have been studied at the intermediate energies $\sqrt{s_ {NN}}<100$ GeV \cite{ref:c,ref:e}. However, insufficient number of particles are produced at those energies to allow bin sizes to be small unless particles produced at all transverse momentum ($\pt$) are included. Such integration over all $\pt$ results in including particles produced at all times, thus smearing out signals of critical behavior and PT that occurs at different times of the evolution in different parts of the medium. At LHC the collision energy is high enough so that it is possible to have $\pt$ cuts and yet still have sufficient number of particles in a small $\pt$ bin to render feasible the study of scaling behavior over a wide range of bin sizes in $(\eta, \phi)$. That is the main reason why we embark on this analysis of investigating the possibility of finding observable signals of quark-hadron PT and the quantitative measure of the critical behavior of the system, since extensive data on global multiplicities at LHC are now available.\\

\par
We develop here the method of analysis to be carried out. Before applying it to the real data we simulate the event structure by use of an event generator, the A Multi-Phase Transport (AMPT) model~\cite{ampt:ref1,ampt:ref1split,ampt:ref2,ampt:ref3}. Recently the AMPT model has successfully 
reproduced many of the experimental data obtained from $\PbPb$ collisions at $\sqrtSnn$ = 2.76 TeV, such as the pseudorapidity and $\pt$ distributions~\cite{ref:amptwork1}, harmonic flows ~\cite{ref:amptwork2,ref:amptwork3} and reconstructed jet observables, including $\gamma$-jet $\pt$ imbalance~\cite{ref:amptwork4}. However, the model does not contain the dynamics of collective interaction, so it is not known  whether the model can generate evidence for critical behavior. Nevertheless, it is a good model for us to use in testing the effectiveness of our method of analysis in search for such evidences in the real data to be analyzed in the future.\\

\par
The organization of this paper is as follows. In Sec.\ II we review the factorial moments, their scaling behaviors and the moments of the event-by-event fluctuations of the spatial patterns. In Sec.\ III the methodology for studying fluctuations in spatial patterns is discussed. A brief introduction to the AMPT model and the simulated data is presented in Sec.\ IV. Results of the analysis are discussed in Sec.\ V followed by summary of the present work in Sec.\ VI. 
%

\section{Scaling Behavior of Factorial Moments}
\subsection{Local Multiplicity Fluctuations of spatial patterns}

 Bialas and Peschanski first proposed the use of factorial moments to study fluctuations because of their property that statistical fluctuations can be filtered out \cite{ref:abialas}. The normalised factorial moment $F_{q}$  is defined as

\begin{equation}
F_{q}(\delta^{d}) = \frac{\langle n!/(n-q)!\rangle}{\langle n \rangle^{q}},
\label{eqfact1}
\end{equation}
where $q$ is the order of the moment and is an integer. $n$ is the number of particles in a bin of size $\delta^{d}$ in a $d$-dimensional space of observables and only $n \ge q$ are counted.  The averages are weighted by the multiplicity distribution $P_n$. It is shown in \cite{ref:abialas} that  if $P_n$ is Poissonian, then $F_q(\delta)=1$ for any $\delta$.  It is in that sense that the statistical fluctuations are filtered out. Thus any increase of $F_q(\delta)$ above 1 implies non-trivial dynamical fluctuations. 
A power-law behavior 

\begin{equation}
F_{q}(\delta) \propto \delta^{-\varphi_{q}}
\label{inter1}
\end{equation}
over a range of small $\delta$ is referred to as  {\em intermittency}, analogous to a similar behavior at the onset of turbulence in fluids for such a dependence. Intermittency implies the lack of any particular spatial scale in the system and has been observed in  many systems of collisions with positive intermittency index ($\varphi_{q}$) \cite{inter:exp}. In terms of the number of bins $M\propto 1/\delta$, thus Eq.\ (\ref{inter1}) may be written as 

\begin{equation}
F_{q}(M) \propto M^{\varphi_{q}},
\label{inter2}
\end{equation}
where $\varphi_q$ is the intermittency index, a positive number. \\
\par
The possibility of finding a signal of quark-gluon plasma by means of intermittency was first pointed out in \cite{bh}. The use of $F_q$ was then applied by Hwa and Nazirov \cite{ref:hn} to the quantification of the nature of fluctuations in a system undergoing second-order PT in the  Ginzburg-Landau (GL) theory~\cite{ref:GL}. It was found that to a high degree of accuracy $F_{q}$ satisfies the power-law behavior 

\begin{equation}
F_{q} \propto F_{2}^{\beta_{q}}, 
\label{fq}
\end{equation}
referred to as F-scaling. Of importance to note is that Eq.\ (\ref{fq}) can be valid even if the scaling behavior in Eq.\ (\ref{inter2}) is not valid. It is derived in \cite{ref:hn} that
\begin{equation}
\beta_{q} = (q-1)^{\nu}, \qquad \nu=1.304 \ .
\label{beta}
\end{equation}
The scaling exponent $\nu$  is essentially independent of the details of the GL parameters. This behavior has been experimentally verified for optical systems at the threshold of lasing \cite{ref:photon}. Theoretically, simulation of quark-hadron PT in the 2D Ising model done in \cite{ref:ising}  results in scaling behavior of $F_{q}$ that is  in agreement with the Eqs.\ (\ref{fq}) and (\ref{beta}). \\
%
%
\subsection{ Moments of Event-by-event Fluctuations of spatial patterns}
 Vertically averaged horizontal moments  can gauge the spatial fluctuations, neglecting the event space fluctuations whereas horizontally averaged vertical moments lose information about spatial fluctuations and only measure the fluctuations from event-to-event. However to fully account for all the fluctuations that a system formed in the heavy ion collisions exhibits, Hwa and Cao \cite{ref:errpap} introduced {\textit{Moments of Factorial Moment distributions}} which takes into account the spatial fluctuations as well as the event space fluctuations. Hwa and Cao have suggested the study of event factorial moments of the spatial patterns, to gauge the degree of event-by-event fluctuations, instead of the sample factorial moments. The analysis is referred to as the {\textit{erraticity analysis}} and the measure of fluctuations of the spatial patterns so determined quantifies this in terms of an index named as {\textit{erraticity index}} ($\mu_{q}$).   In \cite{hy}  it has been proposed to study the erraticity analysis at the LHC energies, to look for the  local mutiplicity fluctuations and hence to quantify these in terms of $\mu_{q}$. It is observed to be a measure,  sensitive to the dynamics of the particle production mechanism and hence to the different classes of quark-hadron phase transition. Below we give a brief introduction to the erraticity analysis.\\
\par
For an $e^{th}$ event, the event factorial moment is  defined as  
\begin{equation}
F_{q}^{e}(M) = \frac{f_{q}^{e}(M)}{[f_{1}^{e}(M)]^{q}},
\label{eq1}
\end{equation}
where  $q$   the order of moment, is a positive integer $\geq 2$  and $M$ is the number of bins. In this equation, the numerator is defined as 
 \begin{equation}
f_{q}^{e}(M)   =  \langle n_{m}(n_{m}-1)......(n_{m}-q+1)\rangle_{h},
 \label{eq2}
\end{equation}
where $n_{m} \ge q$ is the bin multiplicity of the $m^{th}$  bin. $\langle \ldots \rangle_{h}$ defines an average over all bins in one event and is called horizontal average --- in a practice that regards different events as being vertically stacked.  For a two dimensional phase space partitioned into $M^{2}$ bins (each of width $\delta$ on each side) Eq.~(\ref{eq2}) can be rewritten as
\begin{equation}
f_{q}^{e}(M)= \frac{1}{M^{2}} \sum_{m=1}^{M^{2}} n_{m}(n_{m}-1) \ldots (n_{m}-q+1).
\label{eq3}
\end{equation}
The  denominator in Eq.\ (\ref{eq1}) is defined as
\begin{equation}
[f_{1}^{e}(M)]^q= (\frac{1}{M^{2}}\sum_{m=1}^{M^{2}} n_{m})^q.
\label{eq4}
\end{equation}
For every chosen pair of $q$ and $M$, $F^e_q(M)$ is a number characterizing the spatial fluctuations of particles produced in the $e^{th}$ event. Now for a sample of $N$ events, we have a distribution $P(F_{q}^{e}(M))$, since $F_{q}^{e}(M)$ fluctuates from event-to-event.  If vertical average of $F_{q}^{e}(M)$ over all the events is denoted as $\langle F_{q}(M) \rangle_{v}$, then the deviation of $F_{q}(M)$ from $\langle F_{q}(M) \rangle_{v}$ for each event can be given as
\begin{equation}
   \phi_{q}(M) = \frac{F_{q}^{e}(M)}{\langle F_{q}(M)\rangle_{v}}.
    \label{eq5} 
\end{equation}
To quantify the fluctuation of $\phi_q(M)$ from event-to-event one can study (vertical)  moments of the $p^{th}$ 
power 
 of the normalised $q^{th}$ order factorial (horizontal) moments, i.e., $\phi_q(M)$, and define the 
double moment $C_{p,q}$ as
\begin{equation}
C_{p,q}(M) = \langle \phi_{q}^{p}(M)\rangle_{v} = \frac{1}{N}\sum_{e=1}^N [\phi_{q}^{p}(M)]_{e},
\label{eqcpq1}
\end{equation}
where $p$ is a positive real number, not necessarily an integer, and 
\begin{equation}
   \phi_{q}^{p}(M) = \frac{ [F_{q}^{e}(M)]^{p}}{\langle F_{q}(M)\rangle_{v}^{p}}.
   \label{eqphipq}
\end{equation}


To search for $M$-independent property of $C_{p,q}(M)$ one looks for a power-law
 behavior of $C_{p,q}(M)$ in $M$,
\begin{equation}
  C_{p,q}(M) \propto M^{\psi_{q}(p)}.
\end{equation}
If this behavior exists, it is referred to as \textit{erraticity} \cite{ref:errpap}. If $\psi_q(p)$ is found to have a linear dependence on $p$, then  \textit{erraticity index} $\mu_q$ is  defined as 
\begin{equation}
\mu_{q} =\frac{d\psi_{q}(p)}{dp},
\label{muq}
\end{equation}
in the linear region so that it is independent of both $M$ and $p$. Thus $\mu_q$ is a number that characterizes  the fluctuations of spatial patterns.  It is found in \cite{hy}  that $\mu_{4}$ can be an effective measure to distinguish different  criticality classes, viz., critical, quasicritical, pseudocritical and noncritical. For systems with critical transitions $\mu_{4}$ is observed to have low value in comparison to those with random hadronization. $\mu_{q}$ index  gives numerical summary of the different critical cases that is only mildly dependent on $\Delta \pt$. To a good approximation, it is observed~\cite{hy} for the models having contraction owing to confinement, $\mu_{4}$(critical and quasicritical case) = $1.87 \pm 0.84$.  For models without contraction $\mu_{4}$(pseudocritical and noncritical) = $4.65 \pm 0.06$. These are just model values, but suggestive of the significance of erraticity index to characterize dynamical processes.
%
\section{Methodology}
As an effective measure of the fluctuations in the spatial patterns, we find what are known as horizontal factorial moments or the event factorial moments of the multiplicity distributions. 
We have studied charged particles generated in the five $\pt$ windows with $\Delta \pt$ = 0.1 GeV/c for $\pt \le 1.0$ GeV/c so as to avoid smearing of recognizable features, due to superposition of different patterns at different $\Delta \pt$ intervals.\\
\begin{figure}
\begin{minipage}{0.5\textwidth}
\vspace{0.35in}
\centerline{\includegraphics[width=0.95\textwidth,height=2.9in]{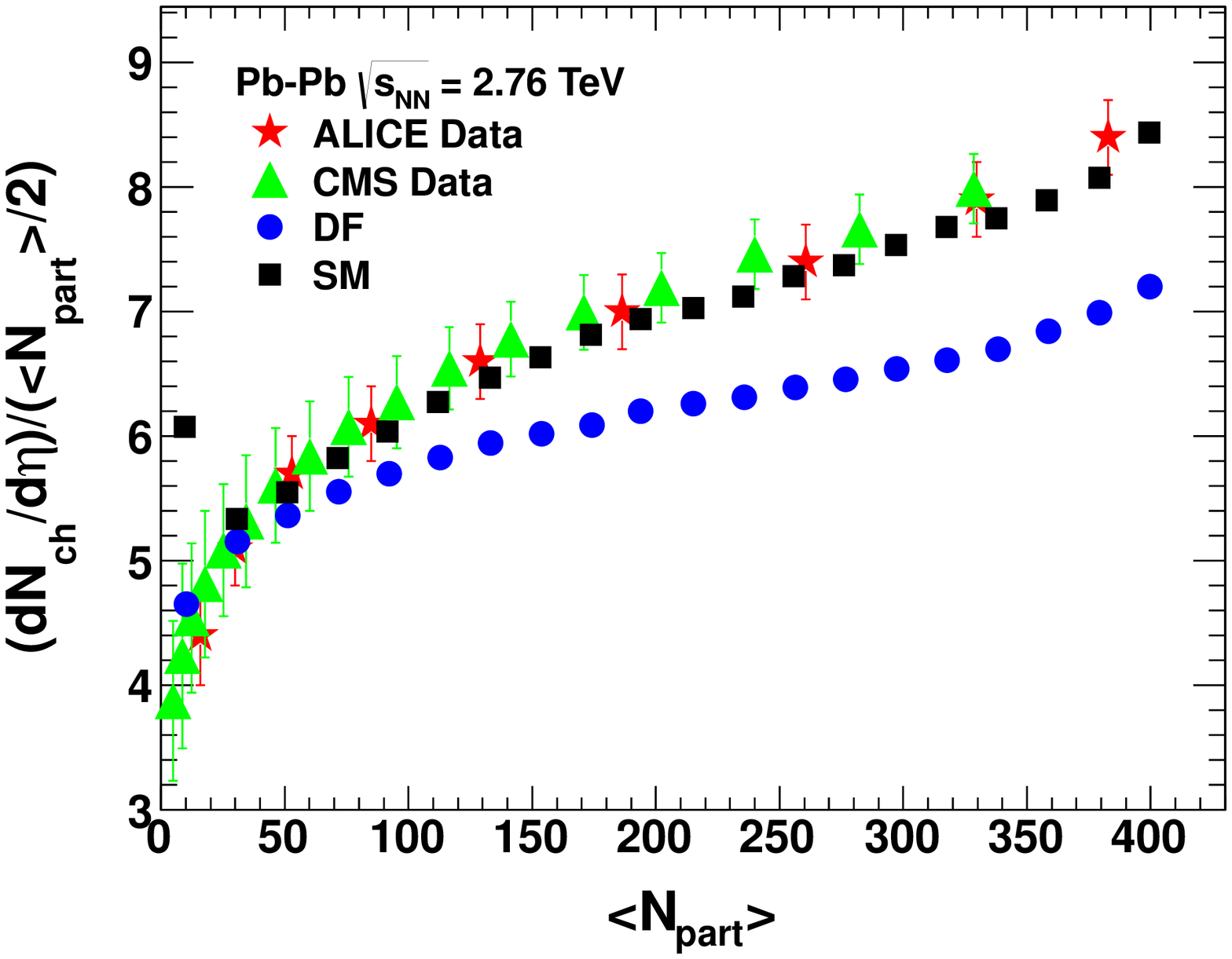}}
\end{minipage}%
\begin{minipage}{0.5\textwidth}
\centerline{\includegraphics[width=0.95\textwidth,height=2.9in]{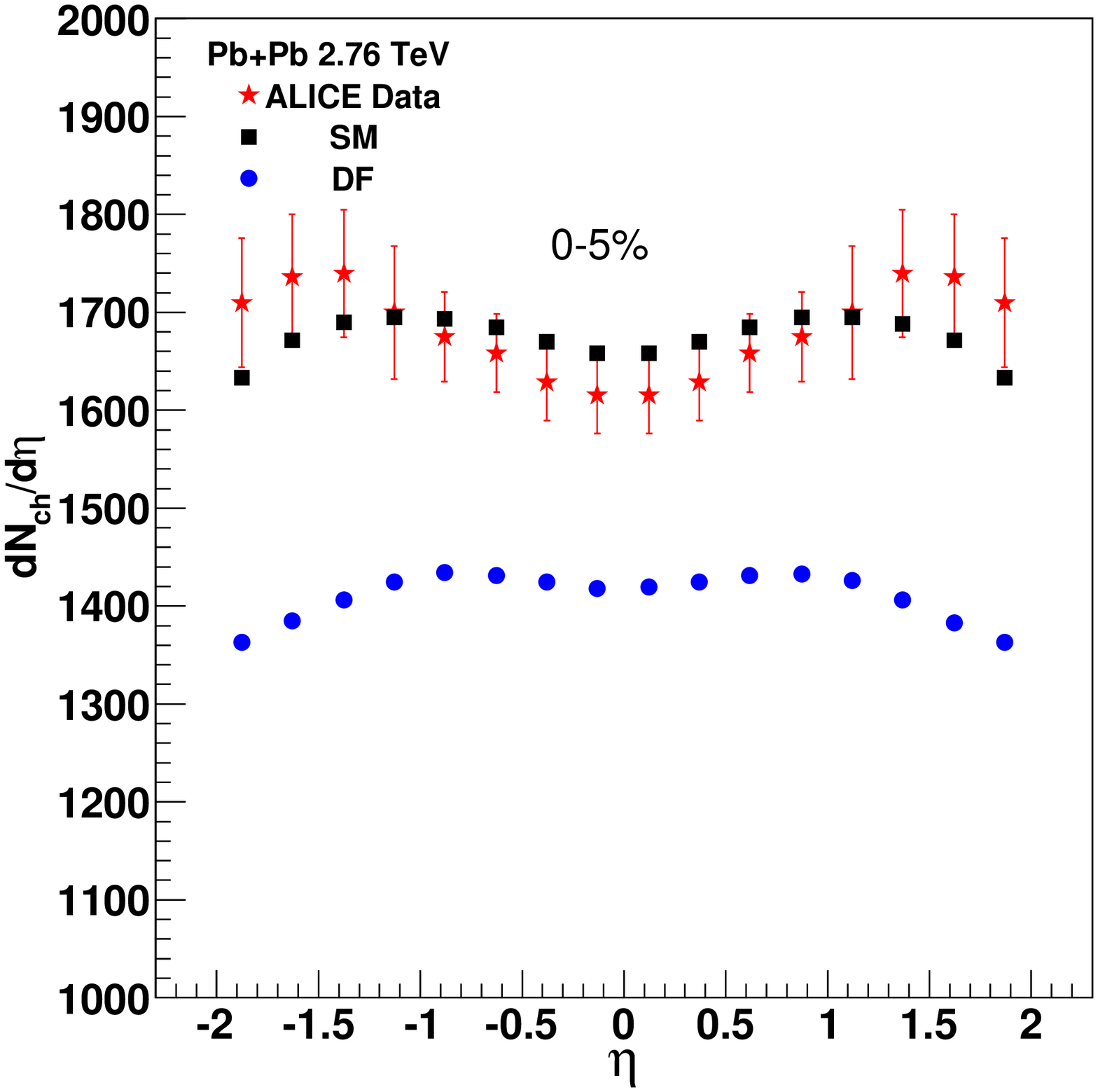}}
\end{minipage}
\caption{(Left) Dependence of $dN_{ch}/d\eta$ on $\langle N_{part} \rangle$ for DF and SM AMPT compared to the ALICE~\cite{fig:alicedata1} and CMS data~\cite{fig:cmsdata}. (Right) Dependence of $dN_{ch}/d\eta$ on $\eta$ for DF and SM AMPT compared to the ALICE Data~\cite{fig:alicedata2}.}
\label{compfig}
\end{figure}     
\par
  Since the single-particle density distribution in pseudorapidity and azimuthal space is non-flat, the shape of this distribution may influence the scaling behavior of the moments. Thus the cumulative variable $X(\eta)$ and $X(\phi)$ are used ~\cite{och:dist} which are defined as 
\begin{equation}
X(y)=\frac{\int_{y_{min}}^{y}\rho(y)dy}{\int_{y_{min}}^{y_{max}}\rho(y)dy}, 
\label{Xdef}
\end{equation}
where $y$ is $\eta$ or $\phi$. Here $y_{min}$ and $y_{max}$ denote respectively the minimum and maximum values of $y$ interval considered, and $\rho(y)$ is the single particle $\eta$ or $\phi$ density. Thus the accessible range of $\eta$ and $\phi$ is mapped to X($\eta$) and X($\phi$) between $0$ and $1$ such that the density of particles is uniform. ($X(\eta),X(\phi)$) unit square of an event in a selected $\pt$ window, is binned into a square matrix with $M^{2}$ bins where the maximum value that $M$ can take depends on the multiplicity in the $\Delta \pt$ interval, so that the important part of the $M$ dependence is captured. Thus for an `$e^{th}$' event, the $q^{th}$ order event factorial moments ($F_{q}^{e}(M)$) as defined in Eq.\ (\ref{eq1}) are determined so as to obtain a simple characterization of the spatial patterns. $F_{q}^{e}(M)$ which is thus a numerical value that describes the pattern of distribution of produced particles of the $e^{\rm th}$ event  are studied for their dependence on $M$ and hence the binning resolution. \\
\begin{figure}[t]
\centerline{\includegraphics[height=4.in,width=0.92\textwidth]{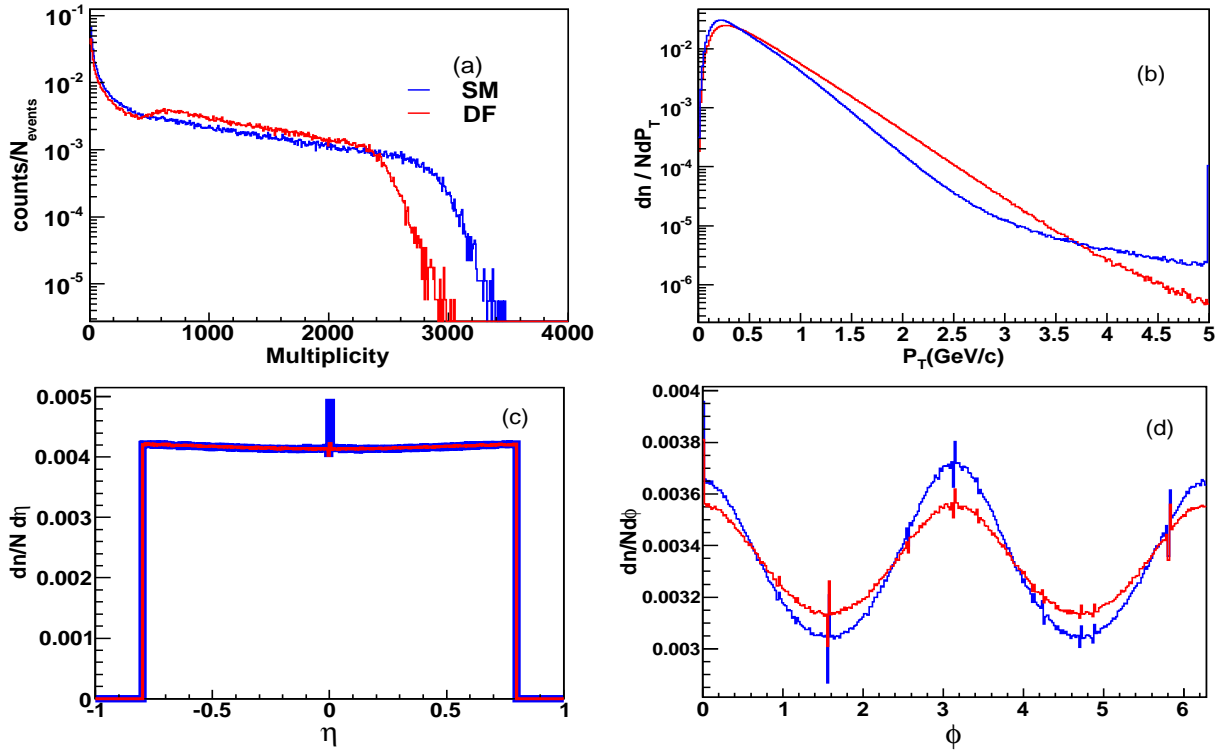}}
\caption{Average particle distributions in AMPT in DF (red) and SM (blue): (a) Distribution in multiplicity $N/N_{event}$ in $|\eta| \le 0.8$, $0 \le \phi \le 2\pi$ and all $\pt$. (b) Distribution in $\pt$. (c)
Distribution in $\eta$. (d) Distribution in $\phi$. }
\label{etaphi}
\end{figure}

\par
If the fluctuations among the bins are Poissonian, then $F_{q}^{e}(M)=1$ for any $M$. If the patterns change from event to event, $F_{q}^{e}(M)$ also changes or fluctuates from event to event and thus one obtains a distribution $P(F_{q}^{e})$ for the whole event sample. Using $\langle F_{q}^{e}(M) \rangle_{v} $ to denote  the (vertical) average of $F_{q}^{e}$ over all events determined from $P(F_{q}^{e})$, one can study  $\langle F_{q}^{e}(M) \rangle_{v} $ or simply  $\langle F_{q} \rangle $ as a function of $M$ and   find out  whether there is intermittency in the data (i.e., Eq.\ (\ref{inter2})) or F-scaling with its associated scaling exponent $\nu$ defined in Eq.\ (\ref{beta}). In the future we will consider higher moments of $F_{q}^{e}(M)$, weighted by $P(F_{q}^{e})$, as suggested in \cite{ch}.\\
\begin{table}
\renewcommand{\arraystretch}{1.5}
\addtolength{\tabcolsep}{2pt}
\centering
\begin{tabular}{c c  c }
\hline  
\bf{$\pt$ window}  & \textbf{Default (DF)}         & \textbf{String Melting (SM)} \\
\textbf{(GeV/c)}         & $<N>$   &  $<N>$       \\
\hline
 $0.2 \le \pt \le 0.3$   & 285.2  & 434.8 \\
 $0.3 \le \pt \le 0.4$   &279.2  & 355.5 \\
 $0.4 \le \pt \le 0.5$   &243.7  & 271.6 \\
 $0.6 \le \pt \le 0.7$   &163.3  & 155.5 \\
 $0.9 \le \pt \le 1.0$   & 80.5  & 66.1\\
\hline
\end{tabular}
\caption{Average Multiplicity of the Simulated Data sets analyzed in different \pt windows.}
\label{t:table1}
\end{table}    
%
\par
To quantify the degree of  fluctuations in the $F_{q}$, we determine the $p^{th}$ power moments of the $F_{q}$, as defined in Eq.\ (\ref{eqcpq1}).
The dependence of the $C_{p,q}$  on $M$ is studied for $q$ = 2, 3, 4 and 5 and $p$ = 1.0, 1.25, 1.50, 1.75 and 2.0. The degree of the event-by-event fluctuations in the spatial patterns, which is quantified by an index $\mu_{q}$, as defined in Eq.\ (\ref{muq}), is determined for the charged particles generated using the two modes of AMPT model. An introduction to the AMPT model and the data generated using it, is given in the next section. \\

\section{A Multi-Phase Transport Model}
The present knowledge of heavy ion collisions demand multi-module modelling to know the detailed description about the entire history of such collisions. The AMPT model has been quite useful in understanding recent experimental results \cite{ampt:ref1,ampt:ref1split,ampt:ref2,ampt:ref3,ref:amptwork1,ref:amptwork2,ref:amptwork3,ref:amptwork4}. A brief introduction of the AMPT model is given here, for details refer to \cite{ampt:ref4, ampt:ref5,ampt:ref6,ampt:ref7}.\\
\begin{figure}[t]
\centerline{\includegraphics[height=2.8in,width=0.8\textwidth]{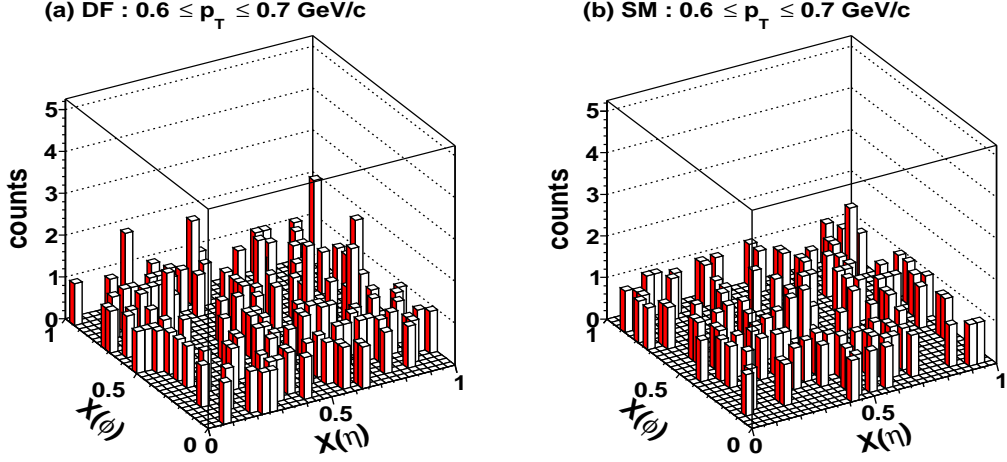}}
\caption{Lego plot of ($X(\eta),X(\phi)$) phase space of an event in the interval $0.6 \le \pt \le 0.7$ GeV/c for $M$ = 32:  (a) DF and   (b) SM.}
\label{lego}
\end{figure} 
\par
The AMPT model is framed to study the nuclear collisions lying in the centre of mass energy range from 5 GeV to 5.5 TeV. It is a hybrid model that includes both initial partonic and the final hadronic state interactions and transition between these two phases and addresses  the non-equilibrium many-body dynamics of  system. In fact, the model consists of four main parts: the initial conditions, partonic interactions, hadronization and hadron rescattering. The initial conditions of spatial and momentum distributions of minijet partons and soft string excitations, for modelling the heavy-ion collisions, are obtained from Heavy Ion Jet INteraction Generator (HIJING) model~\cite{ref:hij}. The subsequent parton-parton elastic scatterings are modelled via the Zhang's Parton Cascade (ZPC) model~\cite{ref:zpc}. There are two modes of the AMPT model $viz$, the Default and the String Melting, depending on how the partons are hadronized. In the Default (DF) mode, which is string and minijet model, minijets partons and strings are produced with the HIJING event generator. The partons are recombined with their parent strings when they stop interacting and the resulting strings are converted to hadrons using the Lund String Fragmentation model. A Relativistic Transport (ART)~\cite{ref:art} model is used to describe how the produced hadrons will interact. In the String Melting (SM) mode of AMPT, the strings produced from HIJING are decomposed into patons which are fed into the parton cascade along with the minijet partons.  A quark coalescence model is used to obtain hadrons from partons and the hadronic interactions are subsequently modelled using ART. It is based on the idea that for energy densities beyond a critical value $1{\rm GeV/fm}^{3}$, it is difficult to visualize the coexistence of strings (or hadrons) and partons. Hence there is need to melt the strings to partons. This is done by converting the mesons to a quark and antiquark pair, baryons to three quarks, etc.  Thus the SM mode includes a fully partonic phase that may be regarded as a QGP, although no thermalization is assumed. It hadronizes through quark coalescence. Studying the events generated by the AMPT model, one can thus investigate systems that may or may not have gone through
 a QGP phase. \\
\par
We have generated events, with  parameters $a=2.2$, $b=0.5$ $\rm GeV^{-2}$, in the HIJING model for Lund string Fragmentation function as used in~\cite{ampt:ref1}. Also the values  $\mu=1.8$ $\rm fm^{-1}$  and $\alpha_{s}= 0.47$, are used for the screening mass in the partonic matter and the strong coupling constant respectively, as used in the AMPT model to describe elliptic flow~\cite{ampt:alphamu1,ampt:alphamu2} and two-pion correlations~\cite{ampt:alphamu3}. These values for the parameters $a$, $b$, $\mu$ and $\alpha_{s}$ were used in the AMPT  for the study of beam energy dependence of anisotropy in the azimuthal distribution ($v_{2}$ and $v_{3}$) of the produced particles ~\cite{dronika:sumit}. Fig.\ \ref{compfig} (left) shows the dependence of charged particle density ($dN_{ch}/d\eta$)  on the participating nucleons ($N_{part}$) and (right) on the $\eta$ for $0-5 \%$ central $\PbPb$ collisions. AMPT DF and SM  data values are compared with the experimental data from LHC. SM mode of AMPT is observed to be in good agreement with the experimental data.  We have generated a total of $23424$  DF and $19669$ SM events with impact parameter  $ \le 5$. 
For charged particles (pions, kaons and protons) with $|\eta| \le 0.8$  and full azimuthal coverage in the $\pt$ windows with width $\Delta \pt = 0.1$ GeV/c,  we have studied
 their local multiplicity fluctuations. Fig.\ \ref{etaphi} shows (a) the multiplicity distributions (normalised with the number of events), (b) $\pt$ distributions ($\le 5$ GeV/c), (c, d) $\eta$ and $\phi$ distributions of the generated data. These $\pt$ bins and respective average charged particle multiplicities are tabulated in Table~\ref{t:table1}.  In those figures we 
 see the differences in the results from the SM and DF settings in AMPT.
 We note that Ref.\ [16] shows a good agreement of the $p_T$ distribution  of the SM version of the AMPT model with that of the LHC data. In the following we move on to observables that have not been analyzed for the LHC data, but we study them first for the simulated data from AMPT.
 \\
\begin{figure}[t]
\centerline{\includegraphics[height=2.8in,width=0.8\textwidth]{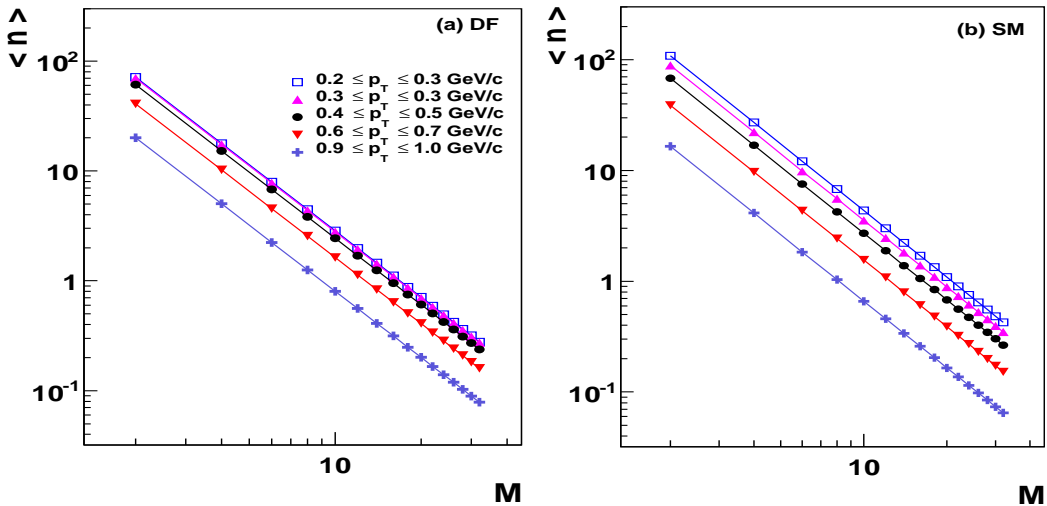}}
\caption{Average charged particle bin multiplicities in the $\pt$ bins  : (a) DF (b) SM AMPT  model.}
\label{figavbin}
\end{figure} 
\section{Results of Analysis}  
From the $X(\eta)$ and $X(\phi)$ distributions obtained from the $\eta$ and $\phi$, as defined in Eq.\ (\ref{Xdef}), we get the $(X(\eta),X(\phi))$ phase space, for each event, which is binned into $M^{2}$ cells  with the minimum value of $M$ being $2$ and the maximum value going anywhere between $10$ to $32$. The maximum $M$ value depends on  the  average bin multiplicity and the order of the moment. Fig.\ \ref{lego} shows a lego plot for some arbitrary DF and SM data event in $0.6 \le \pt \le 0.7$ $\pt$ bin, of the $(X(\eta),X(\phi))$ phase space, having $M$ = 32. We note that there are no bins with $n>2$. That does not mean that there are no events with larger $n$. What we see in Fig.\ \ref{lego} is that there is no obvious clustering.   In Fig.\ \ref{figavbin} the dependence of the average bin multiplicity ($\langle n \rangle$) on $M$  clearly shows the $1/M^{2}$ decrease in log-log plot in  each case of the $p_T$ cuts,
as it should. The point of exhibiting Fig.\ \ref{figavbin} is to show how small $\langle n\rangle$ becomes at high $M$. Furthermore, 
 at low $\pt$, $\langle n \rangle$ is smaller for the DF mode compared  to the SM, but the relative magnitudes are reversed at higher $\pt$.\\
\begin{figure}[h]
\centerline{\includegraphics[height=4.0in,width=0.9\textwidth]{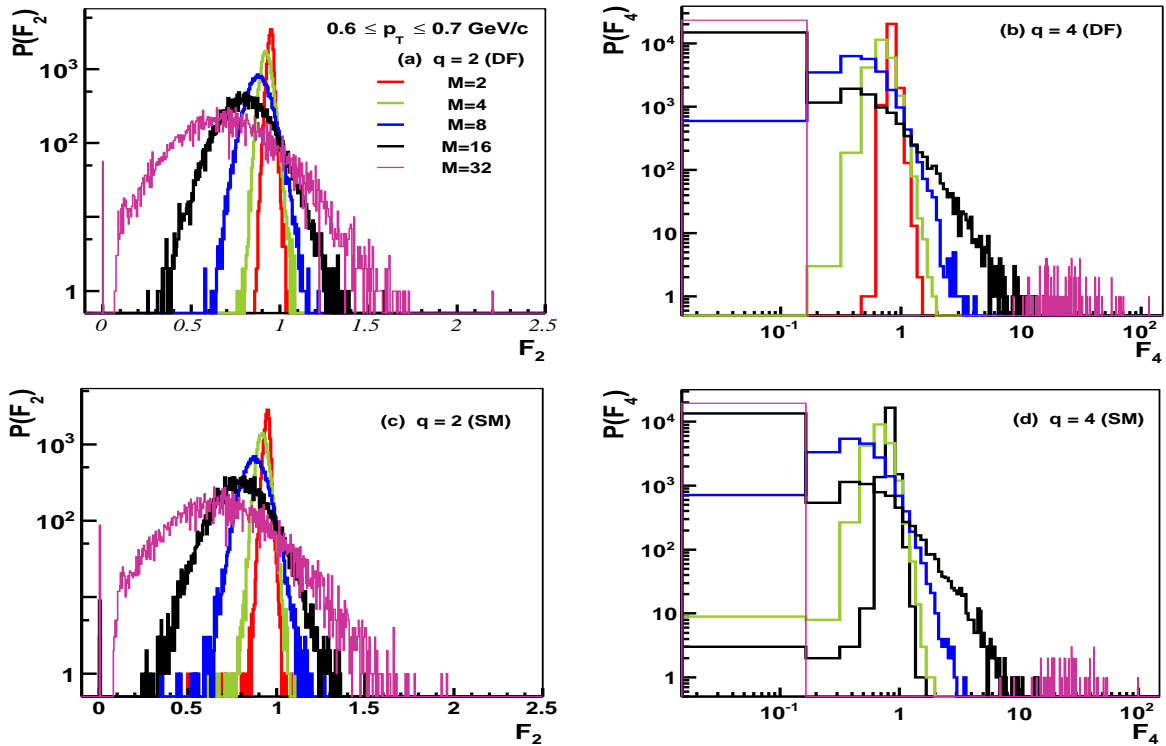}}
\caption{Distributions in $F_{q}$ for $0.6 \le \pt \le 0.7$ GeV/c. (a) q = 2 in DF, (b) q = 4 in DF, (c) q = 2 in SM, (d) q = 4 in SM. }
\label{figfqe}
\end{figure}
\par
Note that at high $M$, $\langle n \rangle$ becomes less than $0.1$ for $\pt>0.9$ GeV/c in both modes. That means that the ($\eta, \phi$) space is very empty. But how the particles in an event under such conditions are distributed over the whole space can fluctuate greatly. Because of the denominator in Eq.\ (\ref{eqfact1}), a cluster of particles with multiplicity $n\ge q$ in an event would produce a large value for $F_q(M)$ for that event. On the other hand, if the particles are evenly distributed, $F_q(M)$ would be smaller. Thus the spatial pattern of the event structure should be revealed in the  distribution of $F_q(M)$ after collecting all events.\\
\begin{figure}[t]
\centerline{\includegraphics[height=3.8in,width=0.8\textwidth]{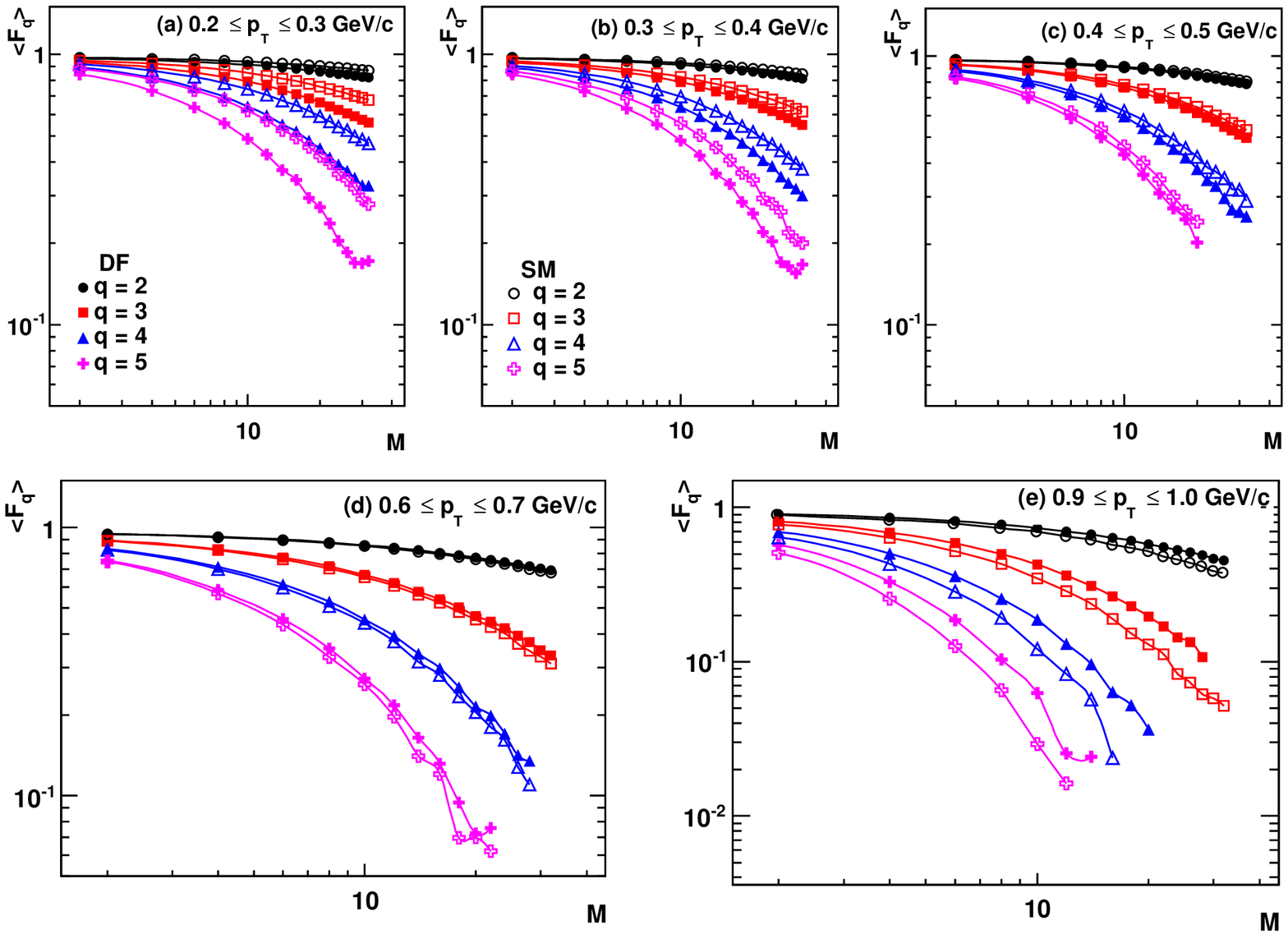}}
\caption{M dependence of $\langle F_{q} \rangle$ in various $\pt$  bins. Solid symbols for DF and open symbols for SM.}
\label{figfqm}
\end{figure}

We have determined the  $P(\fqe)$ distributions  for each data set for $q=2,3,4,5$ and for different $M$ values. We show the $P(\fqe)$ distributions in the $\pt$ window $0.6 \le \pt \le 0.7$ GeV/c in Fig.\ \ref{figfqe}  for $q=2$ in (a) and (c) and for $q=4$ in (b) and (d); the DF and the SM simulated events are in the two rows in the figure. We show only the $M$ values in multiples of 2. Similar distributions are obtained for the other cases also, i.e., for the $q=3$ \& $5$ and for all the $\pt$ bins for both  modes, but are not exhibited here. It is observed that for $q=2$ in (a) and (c), the distributions $F_{2}$ become wider as $M$ increases; however for $q=4$ in (b) and (d) the $F_4$ distributions develop long tails, particularly when $M$ is large. 
Note the log scale on the horizontal axes.
In Fig.\ \ref{figfqe} (a) and (c), for $q$ = 2, the peaks are moving to the left as $M$ increases, thus decreasing $\left< F_{2}\right>$; however, for $q$ = 4, in Fig.\ \ref{figfqe} (b) and (d), the upper tails move towards right  as $M$ increases, while  the lower sides move more and more towards left thereby decreasing the $\langle F_{q}\rangle$.  It means that in small bins the average bin multiplicity $\left< n\right>$ [which is also $f_1(M)$] is so small that when there is a spike of particles in one such bin with $n \ge 4$, the 
non-vanishing numerator in Eq.\ (\ref{eq1}) results in a large value for $F^e_q(M)$ for that $e^{\rm th}$ event. In other words,  for  low average bin multiplicity and at  higher moment-order $q$, there are not many events having $n \ge q$, but when such an event occurs, we have $P(\fqe) \ne 0$ at a high value of $\fqe$. That is the nature of fluctuations in event structure that we look for, as quantified by $P(\fqe)$. As it turns out (to be discussed below), AMPT does not generate enough fluctuations to exhibit properties of PT. \\

\begin{figure}[t]
\centerline{\includegraphics[height=3.8in,width=0.8\textwidth]{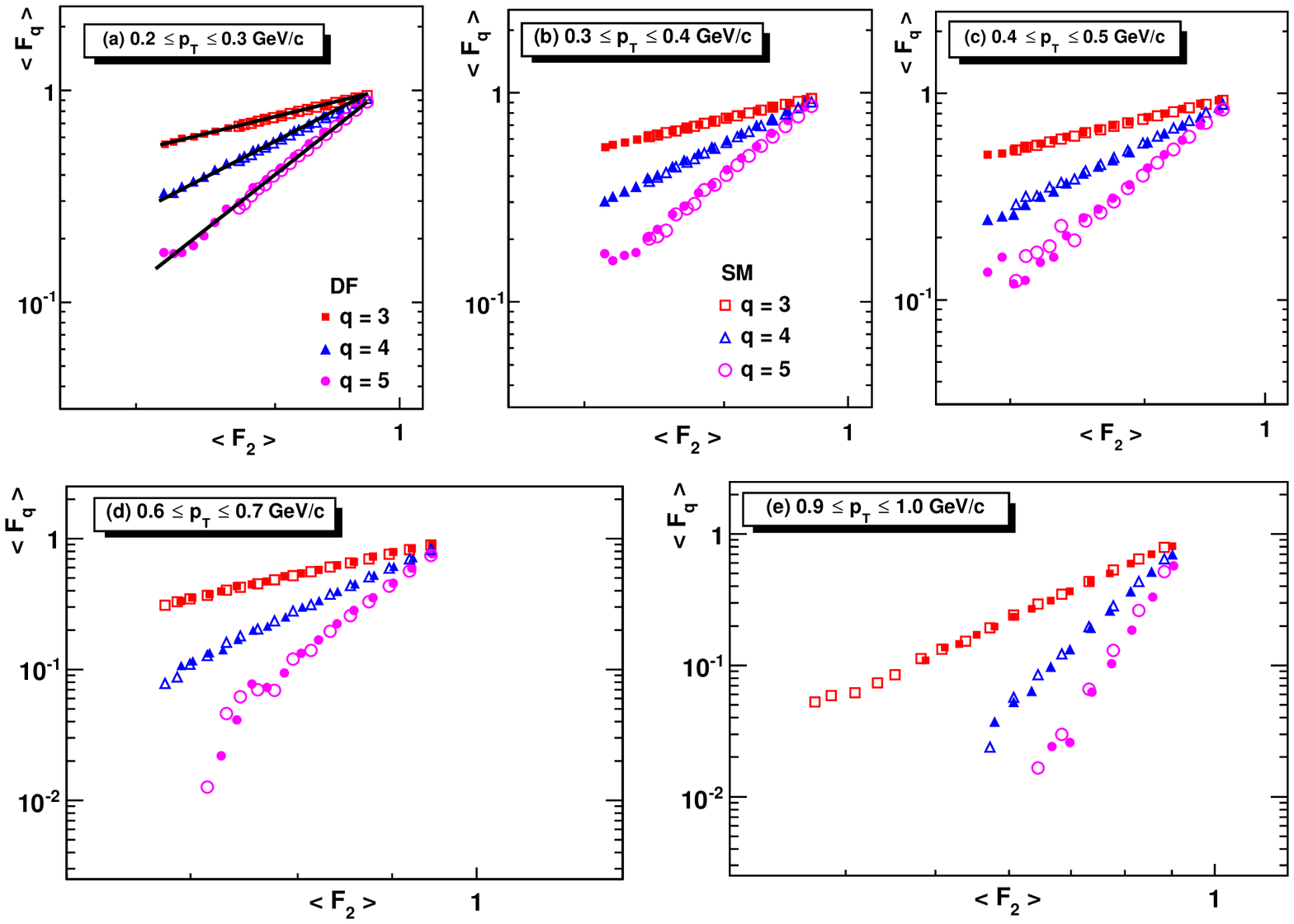}}
\caption{F-scaling plots of $\langle F_{q} \rangle$ vs $\langle F_{2} \rangle$ for various $\pt$ intervals. Solid symbols for DF and open symbols for SM.}
\label{figfqf2}
\end{figure} 
\par
\par
The first moment of $P(\fqe)$ is $ \langle \fqe \rangle$, whose dependence on $M$ can be studied in log-log plots  as  shown in  Fig.\ \ref{figfqm} for various $p_T$ cuts. From the plots, it is observed that values of the moments decrease as the bin size decreases or as $M$ value increases. For both the DF and SM in AMPT  that relationship between $F_q(M)$  and $M$ is  inverse of that in the Eq.\ (\ref{inter2}) such that we represent here the intermittency index as $\varphi_{q}^{-}$ instead of with $\varphi_{q}$; that is

\begin{equation}
F_{q}^{\rm AMPT}(M) \propto M^{\varphi^-_{q}},  \qquad \varphi^-_{q}<0.
\label{interampt}
\end{equation}
%
\begin{figure}
\centerline{\includegraphics[height=2.5in,width=0.8\textwidth]{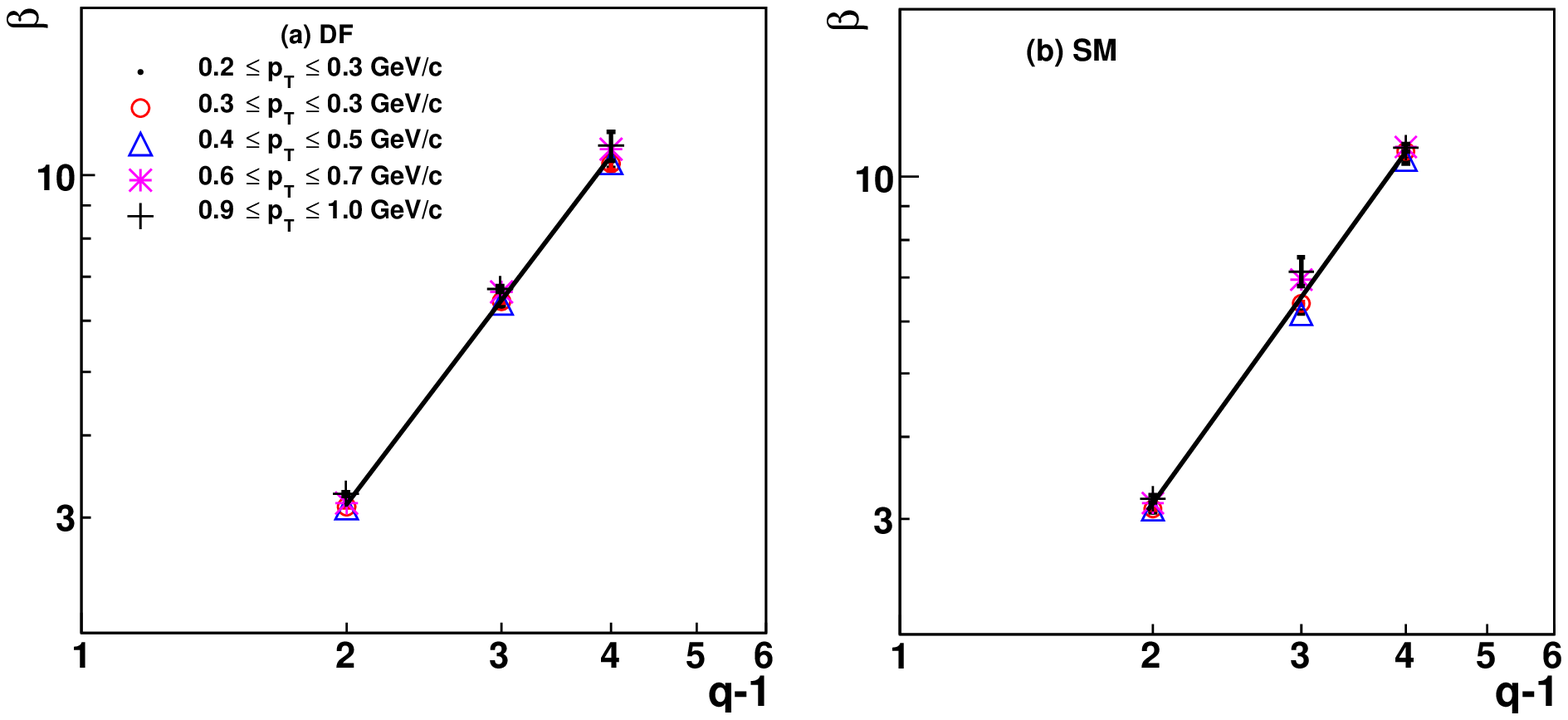}}
\caption{Log-Log plots of $\beta_{q}$ vs $q-1$ for (a) DF and (b) SM. Errors plotted here are the fitting errors,  as obtained from the  line fit of the F-scaling plot.}
\label{betaqall}
\end{figure} 
Hence, with negative $\varphi^-_{q}$ it is found that the charged particles generated by the Default and the String Melting modes of the AMPT model exhibit  {\em inverse of the intermittency behavior}, which we term here as {\em negative intermittency}. Eq.\ (\ref{interampt}) suggests that $F_q^{\rm AMPT}(M)\to 0$ at large $M$ and $q$, implying that the fluctuation is even less than Poissonian. Thus in AMPT there are too few rare high-multiplicity spikes anywhere in phase space.  Eq.\ (\ref{interampt}) is a quantification of the phenomenon exemplified by Fig.\ \ref{lego} for one event, and is a mathematical  characterization after averaging over many events. This same behavior was generated in \cite{hy} for the events belonging to the non-critical class. \\
\par
Though, we observe negative intermittency for both DF and SM mode  of the AMPT data, it is of interest to check whether there is  F-scaling in accordance to Eq.\ (\ref{fq}). We plot  $\fq(M)$  versus $F_{2}(M)$  as is shown in  Fig.~\ref{figfqf2}. Evidently, there is remarkable linearity in the log-log plots, thus revealing the absence of any relevant scale. For each set of points linear fit has been performed to determine the value of the slope, $\beta_{q}$, as exemplified by the straight lines in Fig.\ \ref{figfqf2} (a). Thus we obtain a scaling exponent $\nu_-$ in
\begin{equation}
\beta_{q} = (q-1)^{\nu_{-}},
\label{beta1}
\end{equation}
 similar to that in Eq.\ (\ref{beta}), but for negative intermittency. The dependence of $\beta_q$ on ($q-1$) is shown in Fig.\ \ref{betaqall}, which exhibits good linearity in the log-log plots. The values of  $\nu_{-}$  are  given in Table \ref{t:table2} for different $\pt$ windows and  for both modes of the AMPT model studied here.  Those values (which are positive) should not be compared with  $\nu=1.304$ in Eq.\ (\ref{beta}) because $\nu$ is fundamentally different from $\nu_-$ on account of the difference between the positivity of $\varphi_{q}$ and the negativity of $\varphi^-_{q}$. It is to be noted that the errors on the values are the fitting errors.\\
%
\begin{table}
\renewcommand{\arraystretch}{1.4}
\addtolength{\tabcolsep}{2.5pt}
\centering
\begin{tabular}{c c c}
\hline  
{\bf $\pt$ window}  &      {\textbf {$\nu_{-}$}}             &     {\textbf{$\nu_{-}$}}       \\
 {\textbf(in GeV/c)}           &    {\textbf(DF)}  &  {\textbf(SM)}\\
\hline
 $0.2 \le \pt \le 0.3$   & $1.738 \pm 0.008$  & $1.753 \pm 0.004$  \\
 $0.3 \le \pt \le 0.4$   & $1.774 \pm 0.007$  & $1.793 \pm 0.005$ \\
 $0.4 \le \pt \le 0.5$   & $1.758 \pm 0.006$  & $1.755 \pm 0.006$  \\
 $0.6 \le \pt \le 0.7$   & $1.824 \pm 0.008$  & $1.869 \pm 0.016$ \\
 $0.9 \le \pt \le 1.0$   & $1.778 \pm 0.013$  & $1.781 \pm 0.011$  \\
\hline
\end{tabular}
\caption{Scaling exponents for negative intermittency in the Default and String Melting modes of the AMPT Model.}
\label{t:table2}
\end{table}  
\par
\begin{figure}[t]
\centerline{\includegraphics[height=3.8in,width=0.8\textwidth]{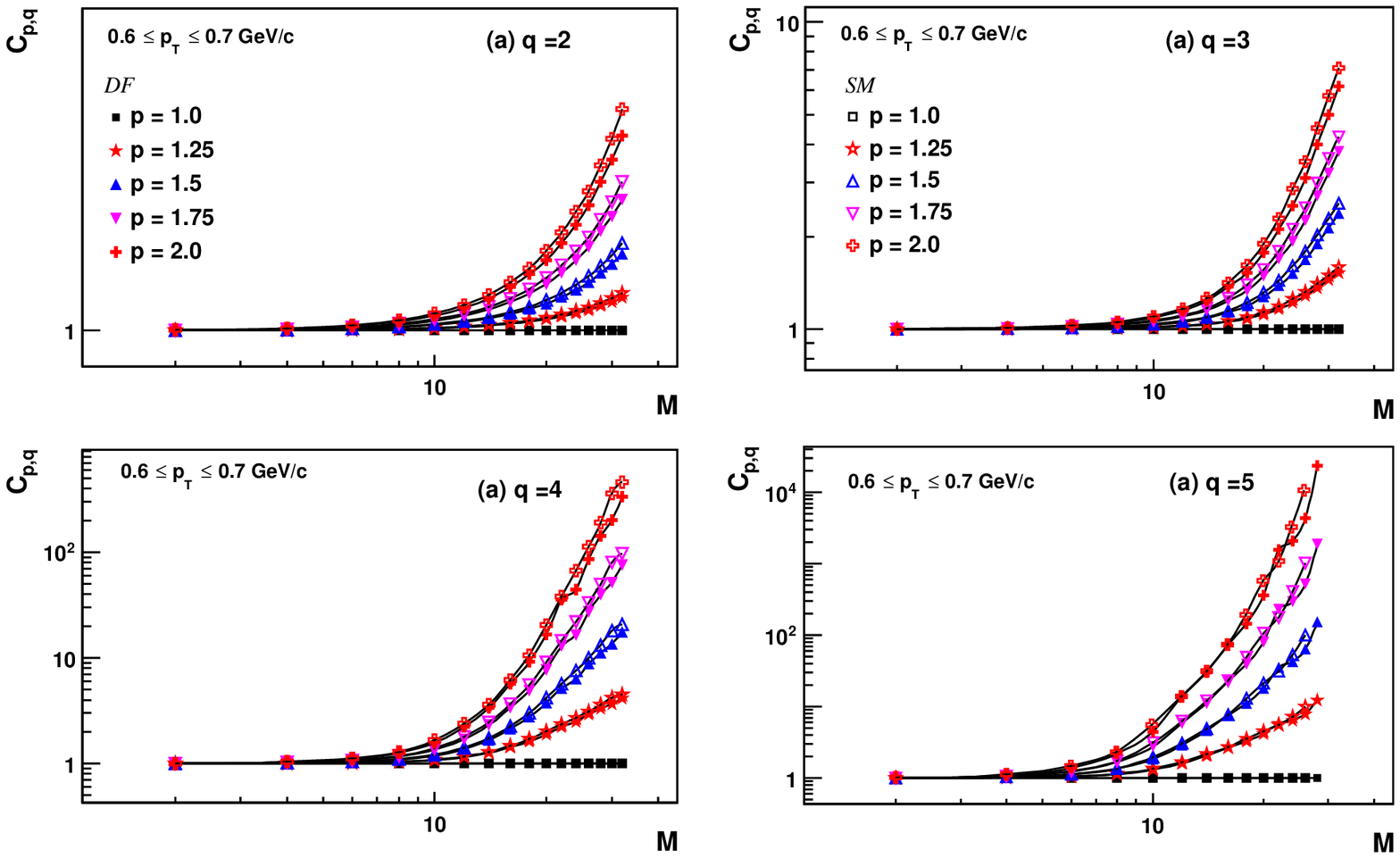}}
\caption{$M$ dependence of $C_{p,q}$ for the $\pt$ window $0.6 \le \pt \le 0.7$ GeV/c in case of DF and SM modes of  the AMPT model.}
\label{cpqM6to7}
\end{figure}
\begin{figure}
\centerline{\includegraphics[height=3.8in,width=0.8\textwidth]{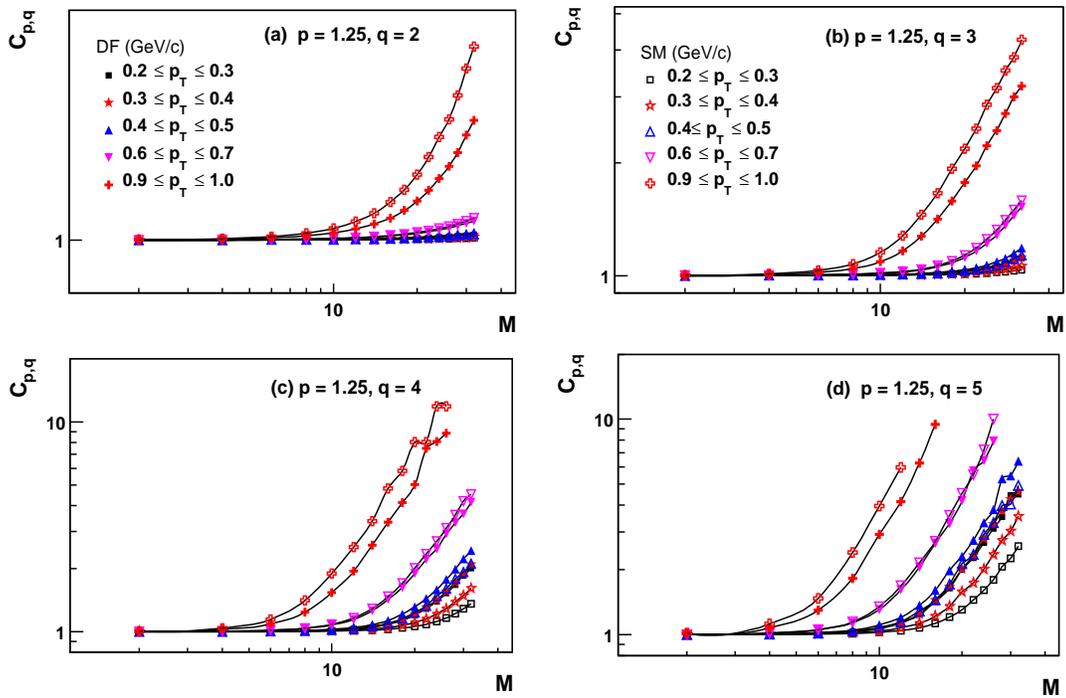}}
\caption{$M$ dependence of $C_{p,q}$ in different  $\pt$ windows for  DF and SM AMPT model, for $p = 1.25$.}
\label{cpqMp125_allwindows}
\end{figure}

 
Since large fluctuations result in the high $F_q$ tails of $P(F_q)$, as exemplified in Fig.\ \ref{figfqe} (b) and (d), it is advantageous to put more weight on the high $F_q$ side in averaging over $P(F_q)$. That is just what the double moment $C_{p,q}(M)$ does. We have determined  $C_{p,q}\rm(M)$ for q = 2, 3, 4, 5 and p = 1.0, 1.25, 1.5, 1.75 and 2.0. The number of bins, $M$, takes on values from 2 to the maximum value  possible while  having reasonable  $\langle n_{m} \rangle$ such that $\fqe \ne 0$. For example for the $0.9 \le \pt \le 1.0$ GeV/c window for q = 5, the average bin multiplicity is very small, so we take $M$ maximum to 12 for DF data. To check whether $C_{p,q}\rm(M)$ follows the scaling behavior with $M$, $C_{p,q}$ is plotted against $M$. Fig.\ \ref{cpqM6to7} (a) to (d) shows respectively, for $q$ = 2, 3, 4 and 5, the $C_{p,q}$ versus $M$ plot in the log-log scale for the window $0.6 \le \pt \le 0.7$ GeV/c,  and for various values of $p$ between 1 and 2. It can be seen that the plots of $\cpq$ versus $M$ have similar shapes for all cases but only with different scales. Further as expected  for  all values of $q$,  for $p$ = 1.0, the $C_{p,q} = 1$ . For $p > 1.0$, $C_{p,q}$ increases with $M$ and $q$ values.  Similar calculations are also done for the other $\pt$ windows as is shown in Fig.\ \ref{cpqMp125_allwindows} (a) to (d) for the case of fixed $p=1.25$  and $q$ = 2, 3, 4, 5 respectively. From the Fig.\ \ref{cpqM6to7} and \ref{cpqMp125_allwindows} we observe that the moment $\cpq$ increases with $M$. Note especially that $C_{p,q}(M)$ in Fig.\  \ref{cpqMp125_allwindows}
  increases as $p_T$ increases. That is due to the fact that at higher $p_T$ the window multiplicity is lower so the bin fluctuation must be larger to register larger $F_q(M)$. \\
\par
\begin{figure}
\centerline{\includegraphics[height=2.8in,width=0.6\textwidth]{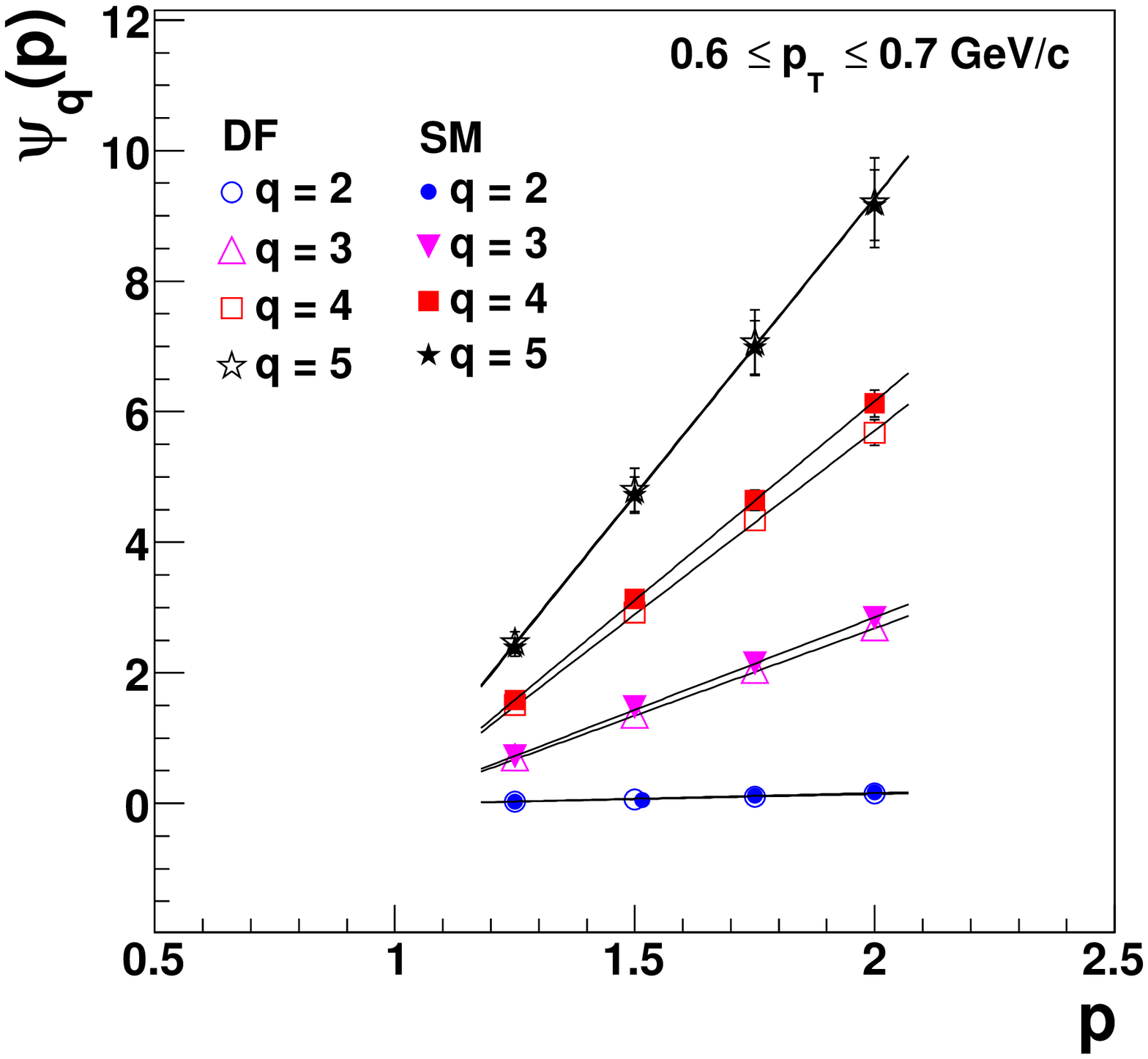}}
\caption{The $\psi_{q}$ dependence on the $p$ for DF and SM AMPT in the $0.6 \le \pt \le 0.7$ GeV/c window. The slope of this gives the $\mu_{q}$.}
\label{psi6to7}
\end{figure}

To extract the erraticity behavior we consider the high $M$ region where 
linear fits are performed  for  each $q$ and $p$ value so as to determine $\psi_{q}(p)$.
We see in Fig.\ \ref{psi6to7}  that for $0.6 \le \pt \le 0.7$ \ $\psi_q(p)$ depends on $p$ linearly for each $q$. Thus the erraticity indices defined in Eq.\ (\ref{muq}) can be determined. Similar plots are obtained for the other $\pt$ windows also. The values of $\mu_q$ with the fitting errors are given in Table \ref{t:table3}.    It can be seen from the table that as $p_T$ value increases (thus decreasing bin multiplicity), the erraticity indices increase for both modes of the AMPT model.  \\
\par
Comparing the $\mu_{q}$ values for the DF and SM data within the same window and for the same values of  $q$, it is observed that $\mu_{q}$ has higher values for the DF mode in comparison to SM for the $\pt$ windows below  $ 0.6 $ GeV/c, but  the values are mixed for $\pt > 0.6$ GeV/c. That phenomenon is related to the average multiplicities of the two modes reversing their relative magnitudes at higher $p_T$, as pointed earlier in connection to Fig.\ \ref{figavbin}.
However it is to be noted from Fig.\ \ref{psi6to7} that the dependence of $\psi_{q}(p)$ on $p$ is distinguishable for the two modes of the AMPT for only $q$ = 4. Coincidentally, as observed in \cite{hy}, $\mu_{4}$ seems to be a good measure to compare the erraticity indices of the different systems and data sets at these energies.\\

\par
We observe that the values of $\mu_{4}^{DF}$ and  $\mu_{4}^{SM}$ for all windows are larger than those obtained for the critical data set in~\cite{hy}, on the same side as the non-critical case. To reflect on the significance of this result, let us point out the difference between erraticity and intermittency. We have found $\varphi^-$ to be negative because $P(F_q)$ broadens, as $M$ increases, with the average $\left< F_q\right>$ shifting to the lower region of $F_q$, thus resulting in negative intermittency. We did notice that the upper tails move to the right, suggesting the presence of some degree of clustering. To emphasize that part of $P(F_q)$ we have taken higher $p$-power moments of $\phi_q(M)$, which suppress the lower side of  $F_q$ while boosting the upper side. The scaling properties of $C_{p,q}(M)$ therefore deemphasize what leads to negative intermittency.  Thus the erraticity indices $\mu_q$ reveal a different aspect of the fluctuation patterns than the scaling indices $\nu^-$. Perhaps they are influenced by the production of jets which show up as clusters, although high-$p_T$ jets are towers in the lego plots in $\eta$-$\phi$ with large $p_T$, while we make low $p_T$ cuts. Details of that aspect of physics require special focused investigation to be done elsewhere. Our study here has revealed interesting properties of scale-invariant fluctuations that should be compared to the real data. Since AMPT contains no dynamics of collective behavior, we should not expect $\nu^-$ and $\mu_q$ to exhibit properties of phase transition.

\section{Summary}
We have studied the local multiplicity fluctuations in the spatial patterns of  charged particles and their event-by-event fluctuations, in central events generated by the Default and the String Melting modes of the AMPT model, using the intermittency and erraticity analysis methodology. We find that the factorial moments decrease with increasing bin numbers, contrary to the usual properties of intermittency observed at lower energies. It means that events with localization of even moderate multiplicities in small bins at low $p_T$ are not generated by AMPT. That is not the property of critical phenomenon, which is supposed to generate fluctuations of all cluster sizes. Since the dynamics of collective behavior is not built into AMPT, no phase transition  of the GL type or   critical behavior discussed in Ref.\ [22]  is expected. The erraticity analysis that we have performed shed further light on the nature of fluctuations, showing that the system generated by AMPT is not near criticality.    The scaling exponent $\nu^-$  and $\mu_{q}$ that we have determined are  useful quantification of our results so that they can be used effectively to compare with other models irrespective of the issues about phase transition. Lastly, the experience that we have gained in this study is extremely beneficial for our analysis of the real data collected at LHC.

\section{Acknowledgements} 
We are thankful to Dr. R.C. Hwa for the very useful discussions and continuous guidance during the analysis of this work. His patience to discuss and timely response to various queries throughout are really appreciated. We are thankful to Prof. S.K. Badyal, Dr. Y.P. Viyogi for the motivation and guidance. We are also thankful to Dr. Tapan K. Nayak and Dr. Premmoy Ghosh of VECC, Kolkata, Dr. Shakeel Ahmad of AMU, Aligarh and Prof. Anju Bhasin of University of Jammu for their inputs on various analysis related details from time to time. We acknowledge the services provided by the grid computing facility at VECC-Kolkata, India for facilitating to perform a part of the computation used in this work. 

\begin{table}
\renewcommand{\arraystretch}{1.2}
\addtolength{\tabcolsep}{2.5pt}
\centering
\begin{tabular}{c  c c c c c}
\hline  
\textbf{q}   & $0.2 \le p_{T} \le 0.3$ &  $0.3 \le p_{T} \le 0.4$ & $0.4\le \pt \le 0.5$   &$0.6 \le \pt \le 0.7$  &   $0.9 \le \pt \le 1.0$          \\
	& & &(in GeV/c) & & \\
\hline
 \textbf{ DF} & & & &  & \\

2    &  $0.043 \pm 0.002$  &  $0.045 \pm 0.002$ & $0.062 \pm 0.003$  & $0.154 \pm 0.008$  & $0.739 \pm 0.043$  \\
3    &  $0.901 \pm 0.081$  &  $0.940 \pm 0.081$ & $1.304 \pm 0.118$  & $2.678 \pm 0.155$  & $4.502  \pm 0.147$ \\
4    &  $4.325 \pm 0.243$  &  $4.532 \pm 0.234$ & $5.478 \pm 0.258$  & $5.640 \pm 0.203$  & $7.484 \pm 0.361$  \\ 
5    &  $6.202 \pm 0.302$  &  $6.150 \pm 0.175$ & $7.396 \pm 0.437$  & $9.107 \pm 0.693$  & $8.643 \pm 0.537$ \\
\hline
 \textbf{SM} & & & & & \\

2    & $0.016 \pm 0.001$  & $0.027 \pm 0.001$  & $0.048 \pm 0.002$   & $0.174 \pm 0.011$  & $1.014 \pm 0.064$  \\
3    & $0.328 \pm 0.025$  &  $0.531 \pm 0.043$ & $1.019 \pm 0.077$   & $2.832 \pm 0.137$  & $ 4.960 \pm 0.150$ \\
4    & $2.481 \pm 0.183$  & $3.385 \pm 0.235$  & $3.935 \pm 0.021$   & $6.101 \pm 0.214$  & $ 7.359 \pm 0.305$ \\
5    & $5.143 \pm 0.022$  & $5.745 \pm 0.312$  & $6.159 \pm 0.280$    & $8.360 \pm 0.533$  & $ 7.655 \pm 0.358$ \\
\hline

\end{tabular}
\caption{Erraticity index in DF and SM modes of the AMPT Model}
\label{t:table3}
\end{table}

\end{document}